\documentclass[useAMS]{mn2e}

\usepackage{graphicx}
\title[The mid-infrared colour-magnitude relation in the Coma cluster]
{The mid-infrared colour-magnitude relation of early-type galaxies in the
Coma cluster as measured by Spitzer-IRS\thanks{This work is
based on observations made with the Spitzer Space Telescope, which
is operated by the JPL, Caltech
under a contract with NASA.}}
\author[Clemens et al.]{M.S. Clemens$^{1}$, A. Bressan$^{1,2,3}$, P. Panuzzo$^{4}$, R. Rampazzo$^{1}$, L. Silva$^{5}$,
\newauthor 
 L. Buson$^{1}$, G.L. Granato$^{5}$\\
$^{1}$INAF-Osservatorio Astronomico di Padova, Vicolo dell'Osservatorio, 5, 35122 Padova, Italy.\\
$^{2}$SISSA-ISAS, International School for Advanced Studies, ia Beirut 4, 34014 Trieste, Italy\\
$^{3}$INAOE, Luis Enrique Erro 1, 72840, Tonantzintla, Puebla, Mexico\\
$^{4}$Laboratoire AIM, CEA/DSM - CNRS - Universit\'{e} Paris Diderot, DAPNIA/Service d'Astrophysique, B\^{a}t. 709, CEA-Saclay,\\ 
F-91191 Gif-sur-Yvette C\'{e}dex, France\\
$^{5}$INAF, Osservatorio Astronomico di Trieste, Via Tiepolo 11, I-34131 Trieste, Italy}

\begin{document}

\date{Accepted ????. Received ????; in original form ????}

\pagerange{\pageref{firstpage}--\pageref{lastpage}} \pubyear{2008}

\maketitle

\label{firstpage}

\begin{abstract}
We use $16\;\rm \mu m$, Spitzer-IRS, blue peakup photometry of 50 early-type 
galaxies in the Coma cluster to define the mid-infrared colour-magnitude 
relation. We compare with recent simple stellar population models that include 
the mid-infrared emission from the extended, dusty envelopes of evolved stars. 
The K$_{\rm s}$-[16] colour in these models is very sensitive to the relative 
population of dusty Asymptotic Giant Branch (AGB) stars. We find that the 
\emph{passively evolving} early-type galaxies define a sequence of approximately 
constant age ($\sim 10$~Gyr) with varying metallicity. Several galaxies
that lie on the optical/near-infrared colour-magnitude relation do not
lie on the mid-infrared relation. This illustrates the sensitivity of the
K${\rm s}$-[16] colour to age. The fact that a colour-magnitude relation is
seen in the mid-infrared underlines the extremely passive nature of the
majority (68 \%) of early-type galaxies in the Coma cluster. The corollary of this is
that 32\% of the early-type galaxies in our sample are \emph{not}
`passive', insofar as they are either significantly younger than 10 Gyr
or they have had some rejuvenation episode within the last few Gyr.
\end{abstract}

\begin{keywords}
galaxies:elliptical and lenticular cD, galaxies:evolution, galaxies:clusters:general, galaxies:photometry.
\end{keywords}

\section{Introduction}

Early-type galaxies (ellipticals and S0s, ETGs hereafter) are ancient stellar
populations whose epoch of formation is related to the large-scale structure formation in the Universe.
This picture is supported by various studies that have used optical
line-strength indices to determine evolutionary parameters of
ETGs in the cluster and field environments (see Renzini 2006 for a review).
Among these studies, some authors (Bernardi et al. 2005,
Clemens et al. 2006, Thomas et al. 2005, S{\'a}nchez-Bl{\'a}zquez
et al. 2006a, Annibali et al. 2007) suggest that cluster ETGs
have a luminosity weighted, mean stellar age 1-2 Gyr older
than those in the field. Recently, Trager, Faber \& Dressler (2008)
challenged this view. Analyzing the Coma cluster, they found that the 12 ETGs
in their sample have mean single stellar population equivalent
ages of 5-8 Gyr, with the oldest systems being $\leq$ 10
Gyr old. This average age is remarkably similar
to the mean age of ETGs in low density environments
(see e.g. Annibali et al. 2007 and references therein).

\begin{figure*}
\centerline{
\includegraphics[scale=0.45]{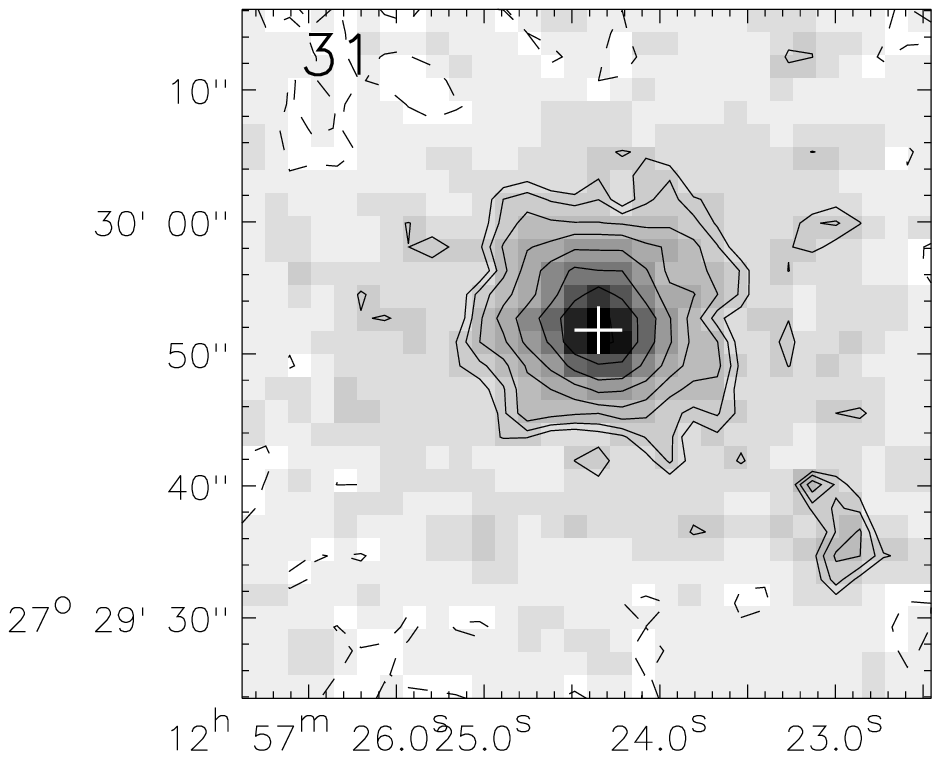}
\hskip-3mm
\includegraphics[scale=0.45]{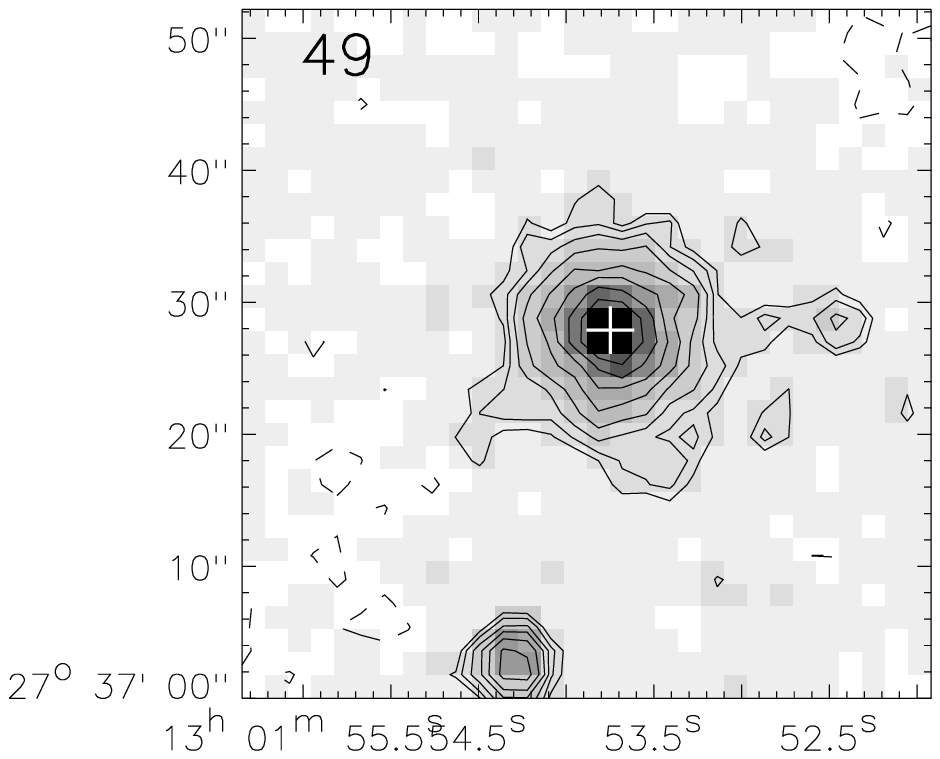}
\hskip-3mm
\includegraphics[scale=0.45]{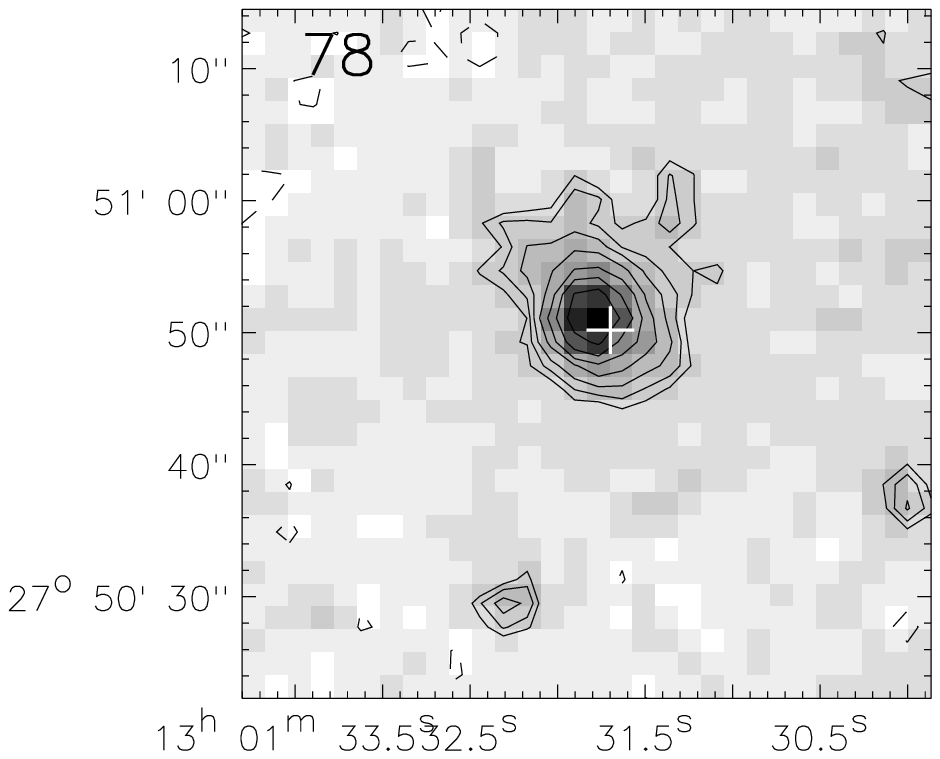}
\hskip-3mm
\includegraphics[scale=0.45]{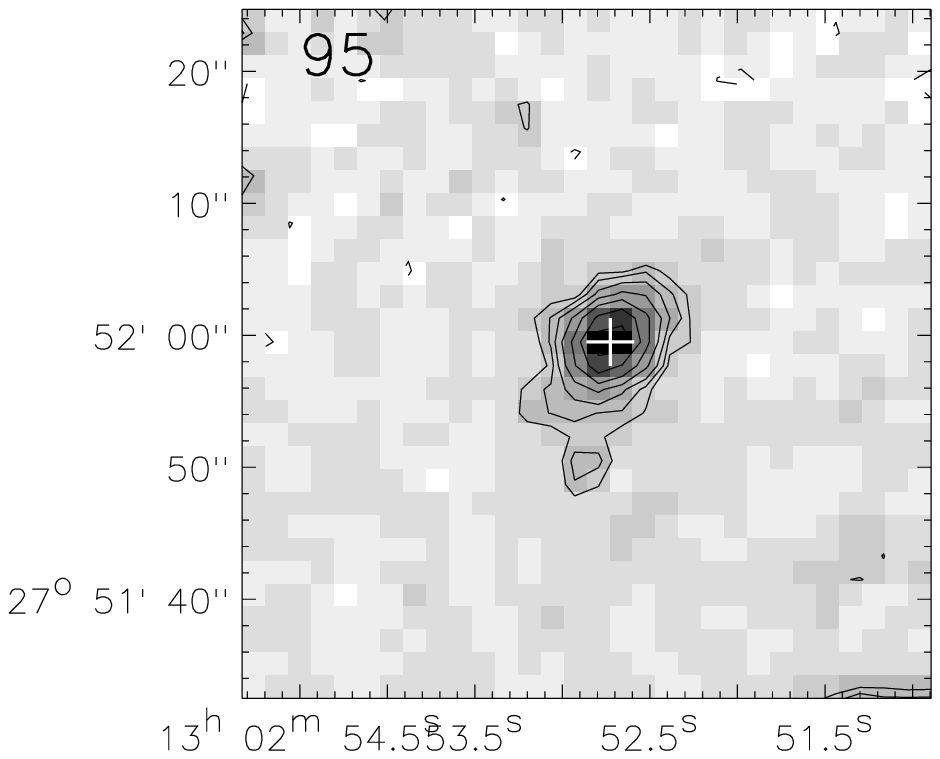}
}
\vskip-5mm
\centerline{
\includegraphics[scale=0.45]{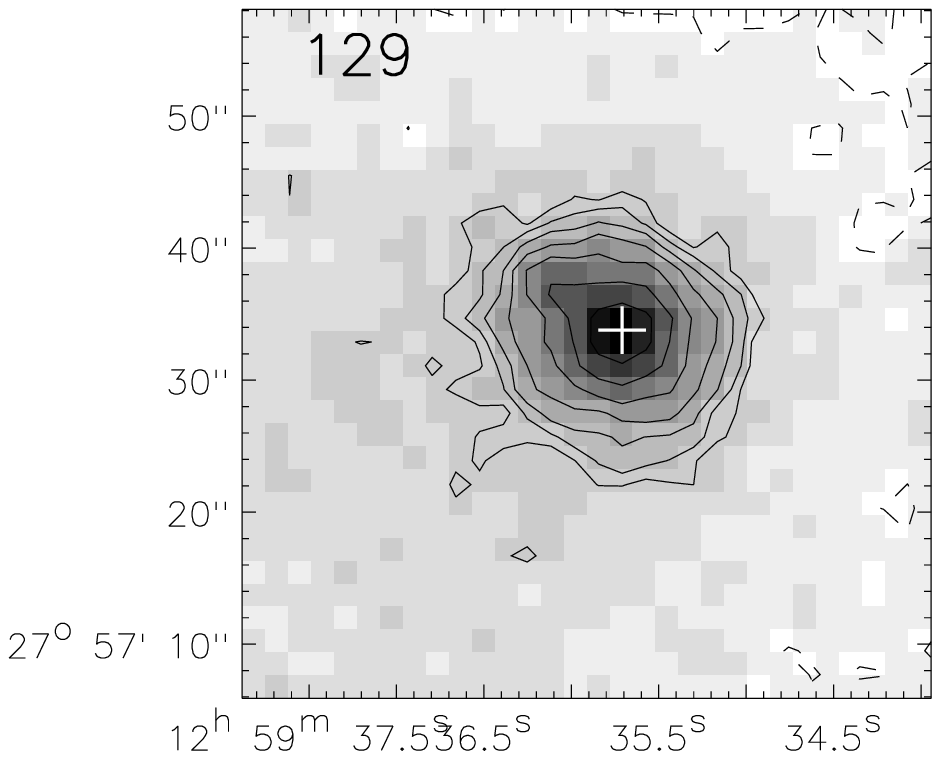}
\hskip-3mm
\includegraphics[scale=0.45]{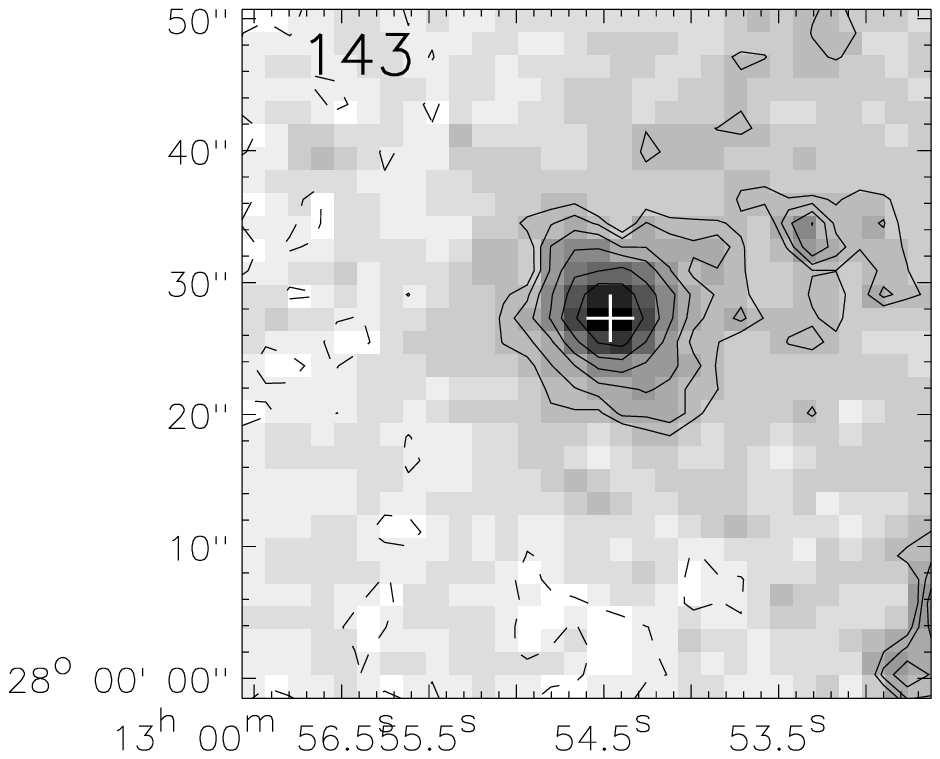}
\hskip-3mm
\includegraphics[scale=0.45]{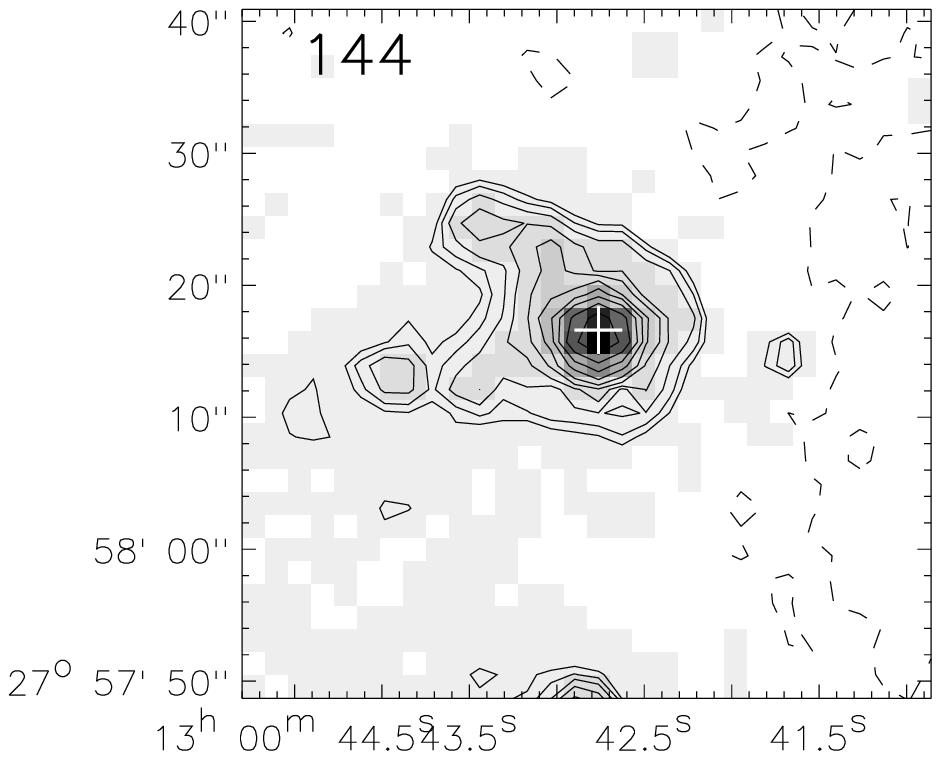}
\hskip-3mm
\includegraphics[scale=0.45]{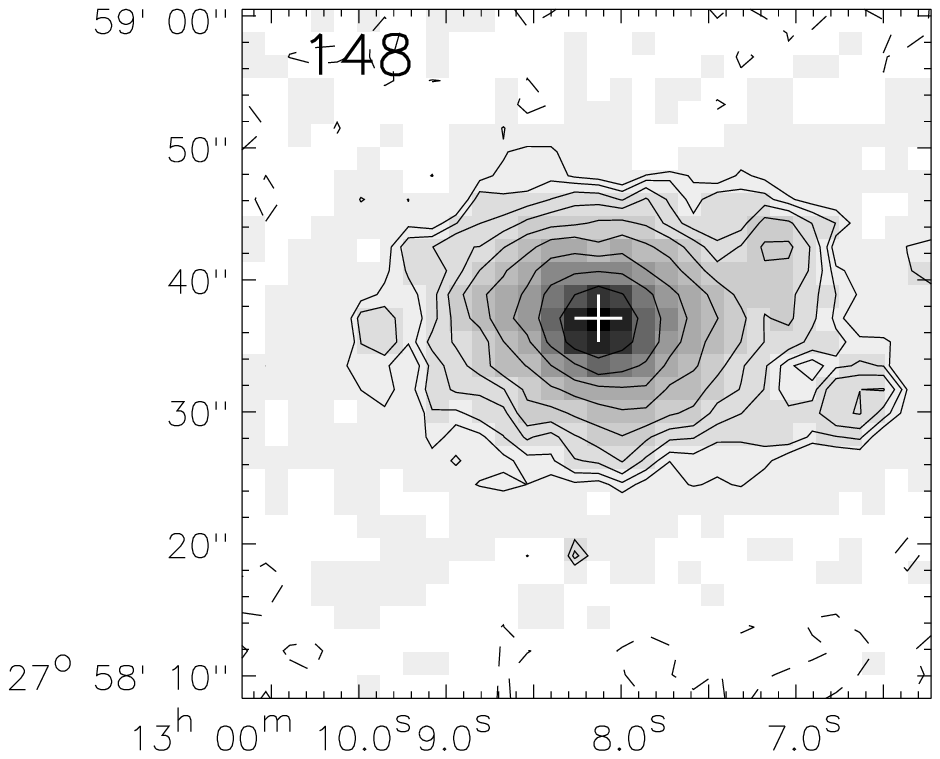}
}
\vskip-5mm
\centerline{
\includegraphics[scale=0.45]{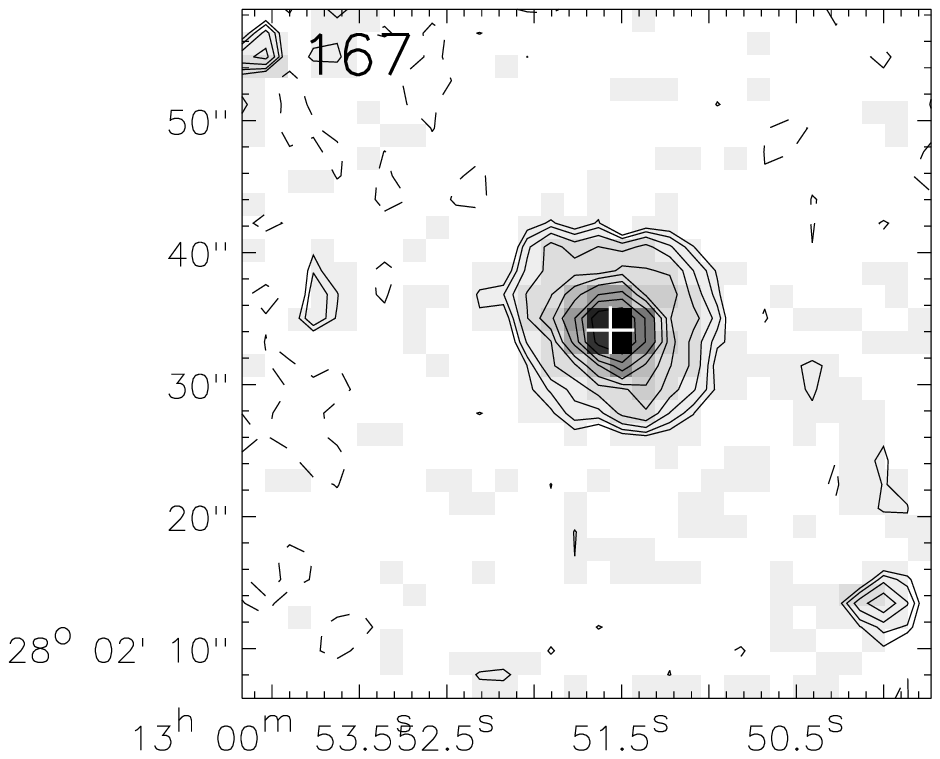}
\hskip-3mm
\includegraphics[scale=0.45]{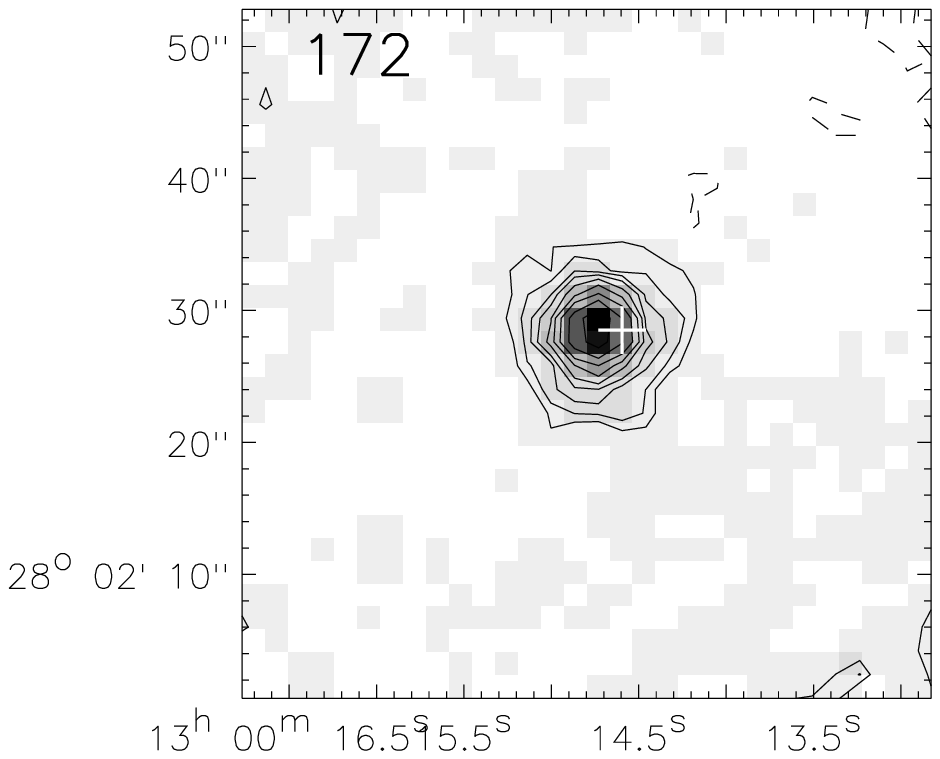}
\hskip-3mm
\includegraphics[scale=0.45]{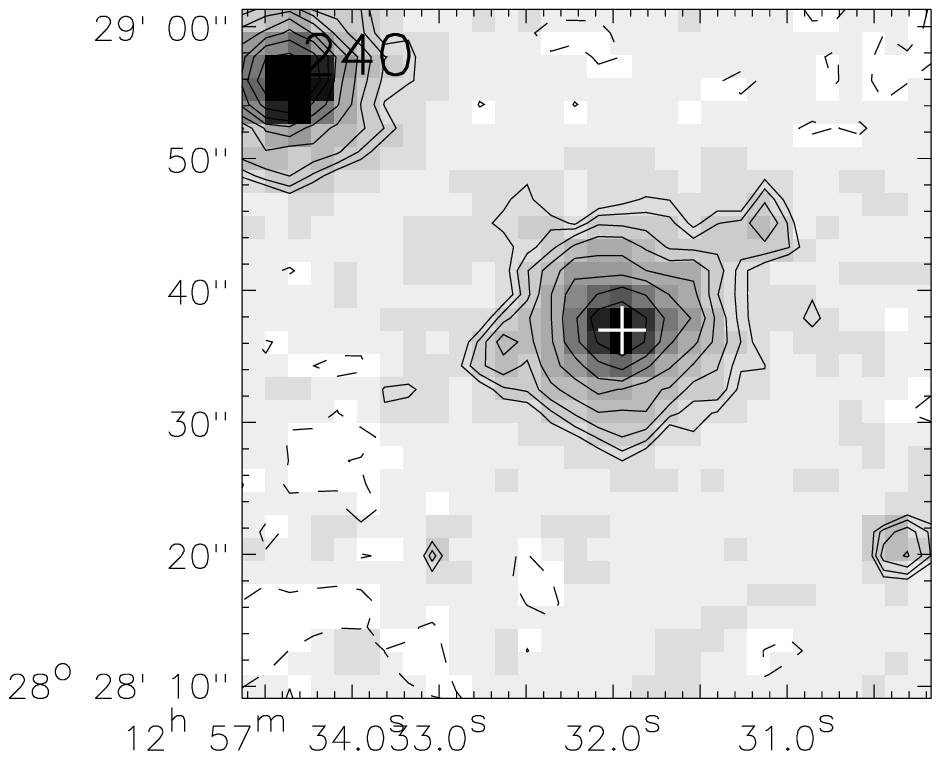}
\hskip-3mm
\includegraphics[scale=0.45]{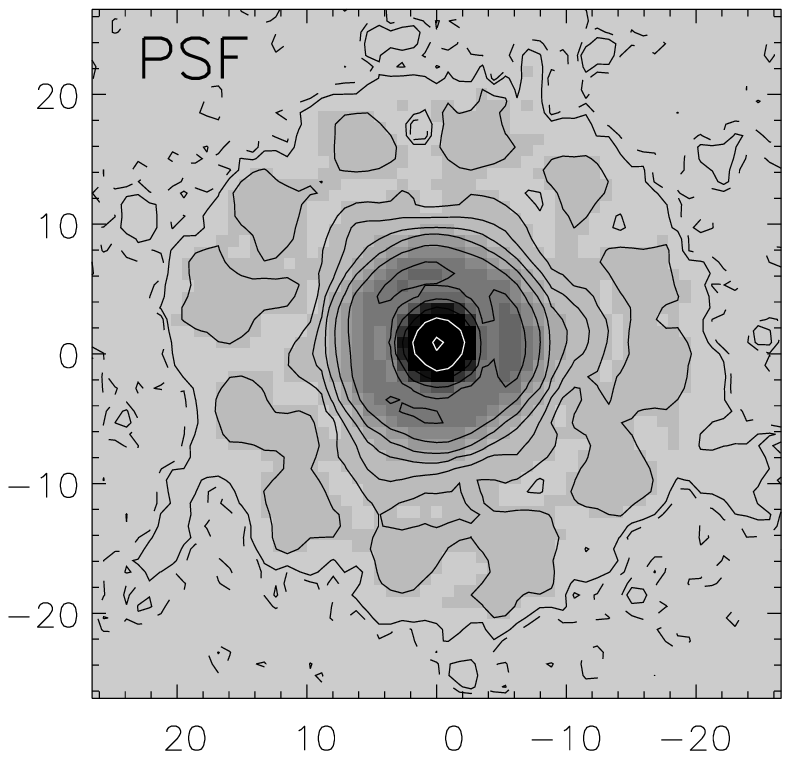}
}
\caption{Example background subtracted, PBCD, IRS peakup images of the Coma ETGs. Contour levels are $2^{n/2}/100$ where n=(-1,0,1,2\dots) $\rm mJy/pixel$ with a dashed contour for negative values at the lowest level. The grey-scale in each plot is scaled to the peak source flux. The last panel is an image of the peak-up point spread function. The cross in each panel gives the optical source position. The grey-scale has a square-root stretch and contour levels are $2^{n/2}/100$ where n=(1,3,5\dots) $\rm mJy/pixel$. Pixels are {1}\farcs{8} on a side.}
\label{fig:maps1}
\end{figure*}

The emission from elliptical galaxies longward of $\sim 10\;\rm \mu m$ has a
large contribution from the circumstellar dust around evolved stars, such as
those on the AGB. This dust emission has been detected by Spitzer in early-type
galaxies where it is seen as a wide emission feature near $10\;\rm \mu m$ with
another broad feature near  $18\;\rm \mu m$ (Bressan et al., 2006). The spatial
profile in the mid-infrared is similar to the optical profile (Temi et al., 2008,
Athey et al. 2002), re-enforcing the notion that the origin is stellar rather
than ISM. This cicumstellar dust has also been detected directly as an extended
envelope around nearby AGB stars (Gledhill \& Yates, 2003). Longer wavelength
infrared emission from elliptical galaxies is sometimes detected (Leeuw et al.,
2004; Marleau et al., 2006; Temi, Brighenti \& Mathews, 2007.), but this cooler
component is probably associated with a diffuse interstellar medium, rather than
with evolved stars directly.

For objects older than $0.1\;\rm Gyr$ the mid-infrared emission traces stars
during their mass-losing AGB phase, and this evolutionary phase is still detected
in objects with ages typical of globular clusters such as 47 Tucanae (Lebzelter
et al., 2006). Because the luminosity of an AGB star depends on its mass and the
main sequence turn-off mass decreases with time, the mid-infrared emission
depends on the age of the stellar population. If all the stars in an early-type
galaxy were created in an instantaneous burst $10\;\rm Gyr$ ago, the mid-infrared
emission would be dominated by stars on the AGB with an initial mass of slightly
less than $1\;\rm M_{\odot}$ (for solar metallicity).

The strength of the silicate emission from the dusty envelopes of evolved stars
is a function of both age and metallicity. As pointed out by Bressan, Granato \& Silva (1998)
however, the dependence of this emission on age and metallicity is different to
that in the optical.

A colour-magnitude relation which includes a band containing the silicate
emission from AGB stars therefore traces the AGB population as a function of the
total luminosity, and can, in principle, be used with the optical relation to
disentangle the effects of age and metallicity.

Here, we use $16\;\rm \mu m$ images of 50 ETGs in the Coma
cluster, made with the blue peakup detector of the IRS on Spitzer. We combine these
with archival Spitzer IRAC images at $4.5\;\rm\mu m$ and K-band images to
construct the mid-infrared colour-magnitude relation. As objects at the
distance of the Coma cluster were too faint for IRS spectroscopy we include 4
galaxies in the Virgo cluster for which we have IRS spectra. These 4 objects were
selected to have spectra that show no evidence of activity, such as recent star
formation or an AGN, and are intended as a `template' for passive objects
against which to compare the more distant objects of Coma.

The Coma cluster is both very rich and dynamically relaxed, and contains some of
the the most massive galaxies in the local Universe. As such, it is an ideal place
to study the star formation history of ETGs.

\begin{figure}
\centerline{
\includegraphics[scale=0.4]{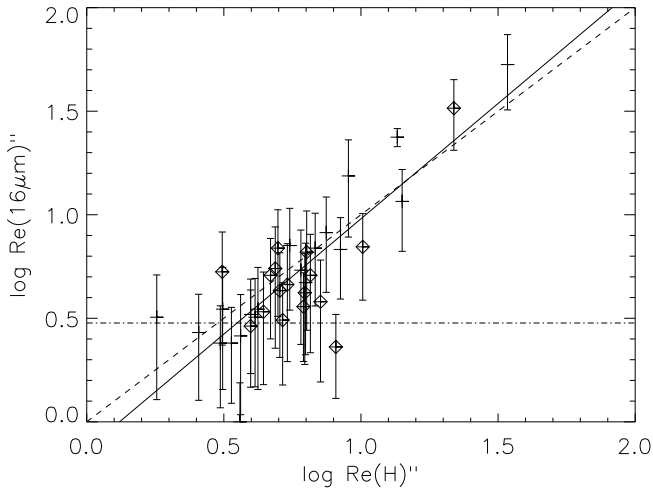}
}
\caption{Relation between the effective H-band radii ($1.65\;\rm \mu m$, Gavazzi et al. 2000) and the $16\;\rm \mu m$ effective radii. The latter have been estimated by convolving a model $r^{1/4}$ de Vaucouleur's law with the IRS blue peakup PSF (see text). Crosses are ellipticals and diamonds are lenticulars (classified either as S0/E, S0, SB0, or S0/a). The solid line is a least squares fit to the data whereas the dotted line corresponds to equal Re at both wavelengths. The horizontal dot-dashed line corresponds to the radius containing half of the encircled energy for the IRS blue peakup PSF.}
\label{fig:radii}
\end{figure}

\section{Data analysis and results}

\subsection{Mid-infrared photometry}

The blue peakup images (see Fig~\ref{fig:maps1} for examples) are background dominated, with source fluxes 
only a few percent above the background level. In addition, the background shows fluctuations on spatial 
scales of several pixels. This presents a
serious problem in the consistent determination of integrated fluxes for sources of different angular sizes. The
standard technique of modeling the background as a smoothly varying light distribution, and then subtracting it
from the image, tends to over-correct in the vicinity of larger sources, simply because the source size is
similar to that of the background fluctuations. In order to avoid this problem we have opted to measure colours,
rather than integrated fluxes, and to do so within a fixed angular radius of $12^{\prime\prime}$. The flux
within this radius is background corrected by subtracting the median pixel value within an annulus just outside
this aperture. The error on the derived flux was taken as the quadrature sum of the rms variation within the
annulus and the calibration error, which was taken as 5\% for the blue peakup data.

Before applying the same procedure to IRAC channel 2\footnote{Calibration errors taken as 10\% , although the
real calibration errors may be smaller.} ($4.5\;\rm \mu m$) and 2MASS Ks-band ($2.2\;\rm \mu m$) images, these
were convolved to the resolution of the IRS-peakup data, in order to remove colour gradients due to the
differing instrumental resolutions. The measured aperture fluxes were then used to derive the colours for each
object; these colours are presented in Table~\ref{tab:fluxes}.

Aperture photometry for the 4 Virgo elliptical galaxies was carried out in an analogous manner, but rather than
use a circular aperture of radius $12^{\prime\prime}$, elliptical apertures of semi-major axis
$72^{\prime\prime}$ were used. This difference takes account of the fact that the Virgo cluster is 6 times
closer than Coma cluster and that the images are well-resolved. There is a possibility that the use of a larger 
aperture in the presence of a fixed psf will introduce a systematic error in the flux determinations 
between the Coma and Virgo galaxies. In order to verify that this was not a significant source of error we 
constructed a number of pairs of simulated galaxy images with de Vaucouleur profiles, differing in their 
effective radii by a factor of 6. We then convolved each with the IRS blue peakup psf and measured the flux 
within apertures of $12^{\prime\prime}$ or $72^{\prime\prime}$ accordingly. The differences in the retrieved 
fluxes did not exceed 10\%, and were typically less than the photometric errors. 

All aperture photometry was performed using a custom pipeline written in {\sc idl} and use was made of 
the {\it IDL Astronomy User's Library} (Landsman 1995).

\subsection{Estimating the effective radii}

A direct estimate of the effective radius based on an image with a high background level and fairly low 
signal-to-noise ratio would be rather sensitive to any error in the subtraction of the background level. 
In order to
minimize this problem, rather than finding the radius containing half of the integrated flux, we adopted the
following procedure. We first constructed a model galaxy with a de Vaucouleur's $r^{1/4}$ law and convolved it
with the IRS blue peakup point spread function (PSF). We either assumed that the intrinsic ellipticity of each
galaxy at $16\;\rm \mu m$ was the same as the optical value as given in ``Goldmine''\footnote{\tt
http://goldmine.mib.infn.it/, Gavazzi et al. (2003)} or was circular, where no
value was available. After convolution with the PSF we measured the radial profile in circular apertures and
compared this to the same apertures measured on the image. The derived value of the effective radius, $R_{e}$,
is that of the model galaxy whose PSF-convolved FWHM was equal to that of the image profile. These values are
given in Table~\ref{tab:fluxes} along with effective radii measured at $1.65\;\rm \mu m$ by Gavazzi et al.
(2000).
The error on the estimate of the effective radii was estimated using a Monte-Carlo approach. To the model
galaxy, convolved with the IRS blue peakup PSF, Gaussian noise was added at the level present in the real
images. For 200 realizations of this noise, the FWHM was measured using the same procedure as for the real
images. The distribution of these values (as a function of S/N ratio) was used as an estimate of the error on
the derived values of Re.

A comparison between the H-band and $16\;\rm \mu m$ effective radii is shown in fig.~\ref{fig:radii}. Clearly,
most galaxies in Coma are barely resolved by Spitzer-IRS, however, for those that are clearly resolved their
effective radii are similar to those in the near-infrared. There is no evidence, for example, that the
$16\;\rm \mu m$ emission is derived from an unresolved central source. Similar effective radii in the near and
mid-infrared are expected if the mid-infrared emission is stellar in origin, such as found by Temi et al. (2008).

\begin{table*}
\centering
\begin{tabular}{c c c c c c c c c c}
\hline\hline
Source ID & GMP83 & R.A. & Dec. & Type & Ks & [4.5]-[16] & $\rm K_{s}-[16]$ & $R_{e}(H)$ & $R_{e}(16\; \rm \mu m)$\\
   & & (J2000) & (J2000) &  & mag &  &  & $^{\prime\prime}$ & $^{\prime\prime}$ \\
\hline
 31 & 4928 & 12h57m24.35s & +27d29m51.8s &  E/S0 &    9.20  &  ---             &  $1.27 \pm 0.056$ & 21.82 & $32.7 \pm 12.2$\\
 49 & 1750 & 13h01m53.75s & +27d37m27.9s &  E    &    ---   &  ---             &  $1.19 \pm 0.056$ &  8.42 & $6.8  \pm 2.9$\\
 67 & 3493 & 12h59m24.93s & +27d44m19.9s &  S0   &    ---   & $0.87 \pm 0.118$ &  $0.95 \pm 0.059$ &  ---  & $3.2  \pm 1.9$\\
 68 & 3660 & 12h59m13.49s & +27d46m28.7s &  S0   &    11.89 & $0.99 \pm 0.117$ &  $1.07 \pm 0.057$ &  7.12 & $3.8  \pm 2.2$\\
 69 & 3730 & 12h59m08.00s & +27d47m02.7s &  E    &    11.09 & $1.19 \pm 0.116$ &  $1.17 \pm 0.056$ &  6.28 & $4.7  \pm 2.6$\\
 70 & 3739 & 12h59m07.51s & +27d46m04.1s &  E    &    11.81 & $0.86 \pm 0.117$ &  $0.90 \pm 0.058$ &  3.13 & $3.5  \pm 2.1$\\
 71 &      & 12h58m57.50s & +27d47m07.5s &  S0   &    12.98 & $0.43 \pm 0.128$ &  $0.55 \pm 0.082$ &  ---  & $2.3  \pm 1.5$\\
 78 & 2000 & 13h01m31.70s & +27d50m50.2s &  E    &    10.72 & $1.11 \pm 0.116$ &  $1.10 \pm 0.056$ &  5.50 & $7.1  \pm 3.6$\\
 79 & 2157 & 13h01m17.50s & +27d48m33.0s &  S0   &    10.85 & $1.26 \pm 0.116$ &  $1.32 \pm 0.056$ &  4.68 & $5.1  \pm 2.6$\\
 84 & 2956 & 13h00m05.30s & +27d48m26.8s &  S0   &    11.87 & $1.54 \pm 0.117$ &  $1.61 \pm 0.056$ &  5.39 & $4.6  \pm 2.7$\\
 87 & 3403 & 12h59m30.64s & +27d47m29.1s &  E    &    12.51 & $0.16 \pm 0.129$ &  $0.18 \pm 0.083$ &  ---  & $1.7  \pm 1.4$\\
 91 & 3997 & 12h58m48.50s & +27d48m37.3s &  S0   &    11.12 & $1.29 \pm 0.116$ &  $1.36 \pm 0.056$ &  5.06 & $4.3  \pm 2.3$\\
 95 &      & 13h02m52.73s & +27d51m59.6s &  S0   &    11.28 &  ---             &  $1.19 \pm 0.056$ &  4.41 & $3.4  \pm 1.9$\\
 96 &      & 13h01m50.22s & +27d53m36.8s &  E    &    11.37 & $1.82 \pm 0.116$ &  $1.94 \pm 0.056$ &  3.62 & $1.0  \pm 0.5$\\
103 & 3400 & 12h59m30.60s & +27d53m03.2s &  S0/a &    11.10 & $1.73 \pm 0.116$ &  $1.78 \pm 0.056$ &  3.96 & $2.9  \pm 1.4$\\
104 & 3296 & 12h59m37.90s & +27d54m26.1s &  S0   &    11.57 & $1.37 \pm 0.116$ &  $1.47 \pm 0.056$ &  3.12 & $5.3  \pm 3.0$\\
110 & 4626 & 12h57m50.62s & +27d52m45.8s &  S0/E &    12.48 & $0.39 \pm 0.11$9 &  $0.39 \pm 0.062$ &  ---  & $2.3  \pm 1.4$\\
118 & 2541 & 13h00m39.50s & +27d55m26.5s &  E    &    11.24 & $1.25 \pm 0.116$ &  $1.32 \pm 0.056$ &  6.04 & $5.4  \pm 3.0$\\
124 & 3201 & 12h59m44.30s & +27d54m44.6s &  E    &    11.34 & $1.40 \pm 0.116$ &  $1.43 \pm 0.056$ &  4.11 & $3.2  \pm 1.7$\\
125 & 3222 & 12h59m42.31s & +27d55m29.1s &  E    &    12.57 & $0.85 \pm 0.120$ &  $0.92 \pm 0.063$ &  1.80 & $3.2  \pm 1.9$\\
126 &      & 12h59m44.00s & +27d57m30.3s &  S0   &    12.69 & $0.71 \pm 0.120$ &  $0.76 \pm 0.064$ &  ---  & $5.5  \pm 2.6$\\
129 & 3329 & 12h59m35.71s & +27d57m33.8s &  D    &    8.86  & $1.26 \pm 0.116$ &  $1.29 \pm 0.056$ & 34.23 & $53.1 \pm 21.0$\\
132 & 3487 & 12h59m25.31s & +27d58m04.7s &  S0   &    12.61 & $0.98 \pm 0.118$ &  $1.09 \pm 0.059$ &  ---  & $3.5  \pm 2.1$\\
133 & 3639 & 12h59m15.00s & +27d58m14.9s &  E    &    11.00 & $1.87 \pm 0.116$ &  $1.84 \pm 0.056$ &  3.37 & $2.4  \pm 1.2$\\
134 &      & 12h59m03.85s & +27d57m32.6s &  E    &    13.13 & $0.92 \pm 0.120$ &  $0.84 \pm 0.065$ &  ---  & $2.2  \pm 1.2$\\
135 & 3851 & 12h58m59.86s & +27d58m02.6s &  E    &    12.89 & $0.54 \pm 0.122$ &  $0.90 \pm 0.068$ &  ---  & $3.9  \pm 1.8$\\
136 & 3914 & 12h58m55.30s & +27d57m52.5s &  E    &    12.52 & $0.73 \pm 0.117$ &  $0.76 \pm 0.059$ &  ---  & $2.0  \pm 0.9$\\
143 & 2390 & 13h00m54.46s & +28d00m27.3s &  E    &    10.41 & $1.07 \pm 0.116$ &  $1.10 \pm 0.056$ &  9.00 & $15.4 \pm 7.6$\\
144 & 2516 & 13h00m42.76s & +27d58m16.6s &  S0/a &    10.76 & $2.10 \pm 0.116$ &  $2.17 \pm 0.056$ &  6.17 & $3.6  \pm 1.6$\\
145 & 2535 & 13h00m40.70s & +27d59m47.9s &  S0   &    11.67 & $0.92 \pm 0.118$ &  $0.99 \pm 0.059$ &  4.87 & $5.5  \pm 3.2$\\
148 & 2921 & 13h00m08.13s & +27d58m37.1s &  D    &    8.41  & $1.28 \pm 0.116$ &  $1.31 \pm 0.056$ & 13.57 & $23.7 \pm 2.4$\\
150 & 2940 & 13h00m06.20s & +28d00m14.7s &  E    &    12.14 & $1.49 \pm 0.116$ &  $1.62 \pm 0.056$ &  4.21 & $3.5  \pm 2.1$\\
152 & 3170 & 12h59m46.79s & +27d58m26.0s &  SB0  &    11.57 & $1.18 \pm 0.119$ &  $1.27 \pm 0.063$ &  6.34 & $6.6  \pm 3.8$\\
155 & 3367 & 12h59m32.82s & +27d59m01.2s &  S0   &    11.25 & $1.66 \pm 0.116$ &  $1.75 \pm 0.056$ &  5.18 & $3.1  \pm 1.6$\\
157 & 3484 & 12h59m25.46s & +27d58m23.7s &  S0   &    12.45 & $1.26 \pm 0.117$ &  $1.37 \pm 0.058$ &  ---  & $10.9 \pm 4.9$\\
160 & 3761 & 12h59m05.90s & +27d59m48.2s &  SB0  &    11.38 & $1.64 \pm 0.117$ &  $1.77 \pm 0.058$ &  6.24 & $4.2  \pm 2.3$\\
164 &      & 13h03m00.89s & +28d01m57.2s &  S0   &    10.31 &  ---             &  $1.22 \pm 0.056$ & 10.15 & $7.0  \pm 3.1$\\
167 & 2417 & 13h00m51.57s & +28d02m34.2s &  S0/E &    10.66 & $1.88 \pm 0.116$ &  $1.94 \pm 0.056$ &  8.10 & $2.3  \pm 1.0$\\
168 & 2440 & 13h00m48.67s & +28d05m26.9s &  E    &    10.90 & $1.58 \pm 0.116$ &  $1.68 \pm 0.056$ &  3.96 & $3.3  \pm 1.6$\\
170 & 2727 & 13h00m22.00s & +28d02m50.1s &  SB0  &    11.55 & $1.22 \pm 0.117$ &  $1.34 \pm 0.056$ &  6.54 & $5.1  \pm 2.9$\\
171 &      & 13h00m16.90s & +28d03m50.0s &  S0   &    12.40 & $0.74 \pm 0.122$ &  $0.72 \pm 0.069$ &  ---  & $3.6  \pm 2.2$\\
172 & 2839 & 13h00m14.60s & +28d02m28.6s &  E    &    11.78 & $1.98 \pm 0.116$ &  $2.06 \pm 0.056$ &  3.07 & $2.4  \pm 1.2$\\
174 & 2922 & 13h00m07.80s & +28d04m42.7s &  E    &    11.59 & $1.66 \pm 0.116$ &  $1.73 \pm 0.056$ &  2.56 & $2.7  \pm 1.4$\\
175 & 3073 & 12h59m55.90s & +28d02m04.9s &  S0   &    11.16 & $1.52 \pm 0.117$ &  $1.65 \pm 0.057$ &  4.98 & $6.9  \pm 3.7$\\
193 & 3084 & 12h59m55.10s & +28d07m42.2s &  E    &    12.28 & $0.87 \pm 0.117$ &  $0.92 \pm 0.058$ &  ---  & $3.8  \pm 2.3$\\
194 & 3792 & 12h59m03.79s & +28d07m25.6s &  E    &    10.29 & $1.13 \pm 0.117$ &  $1.15 \pm 0.057$ &  6.81 & $6.9  \pm 3.3$\\
207 & 2912 & 13h00m09.14s & +28d10m13.6s &  E    &    11.78 & $1.20 \pm 0.118$ &  $1.32 \pm 0.059$ &  3.64 & $2.6  \pm 1.5$\\
217 & 3055 & 12h59m57.60s & +28d14m50.6s &  E    &    10.39 & $1.23 \pm 0.116$ &  $1.29 \pm 0.056$ &  7.47 & $8.2  \pm 4.0$\\
240 & 4822 & 12h57m31.95s & +28d28m37.0s &  E    &    9.22  &  ---             &  $1.33 \pm 0.056$ & 14.13 & $11.6 \pm 4.9$\\
245 &      & 12h56m56.58s & +28d37m24.1s &  S0   &    12.05 &  ---             &  $1.29 \pm 0.057$ &  ---  & $3.2  \pm 1.6$\\
\hline
\end{tabular}
\caption{The integrated $16\; \rm \mu m$ fluxes and sizes of the sample ETGs. The flux errors are the quadrature sum of a term due to the uncertainty in deriving the background sky level and a 5\% calibration uncertainty. The effective radii have been calculated by convolving a model $r^{1/4}$ de Vaucouleur's law with the IRS blue peakup PSF (see text).}
\label{tab:fluxes}
\end{table*}

\begin{figure}
\includegraphics[clip,width=8.3cm]{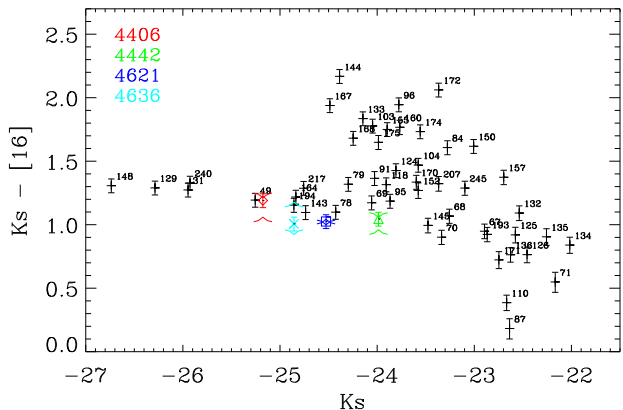}
 \caption{Mid-infrared colour-magnitude relation of the Coma cluster
 for fixed aperture K$_{\rm s}$-[16] colours. 
 Objects are identified according to their designations in Dressler (1980). 
 The 4 ETGs in the Virgo cluster
 are shown with coloured symbols. For each of the Virgo galaxies the mid-infrared
 colours within the central $10^{\prime\prime}$ radius, as well as that of the
 whole galaxy \emph{excluding} this central aperture are shown, as upward and 
downward pointing symbols, respectively. 
 The assumed distance moduli to the Coma and
 Virgo clusters are 35.1 and 31.3 respectively.}
\label{fig:cm12}
\end{figure}

\begin{figure}
\includegraphics[clip,width=8.3cm]{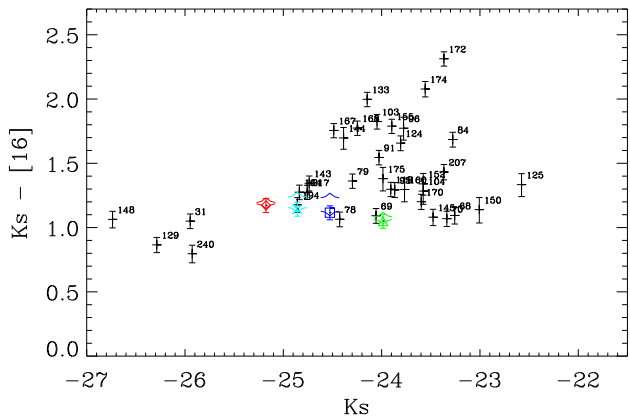}
 \caption{As fig.~\ref{fig:cm12} but for photometry with circular radii equal to 
$2R_{\rm e}$ where $R_{\rm e}$ is the effective radius of the galaxy measured in 
the H-band (Gavazzi et al., 2000). For the 4 Virgo galaxies a radius of only 
$R_{\rm e}$ was used due to the limited size of the peakup images.}
\label{fig:re2}
\end{figure}

\begin{figure}
\centering
\includegraphics[clip,width=5.3cm,angle=90]{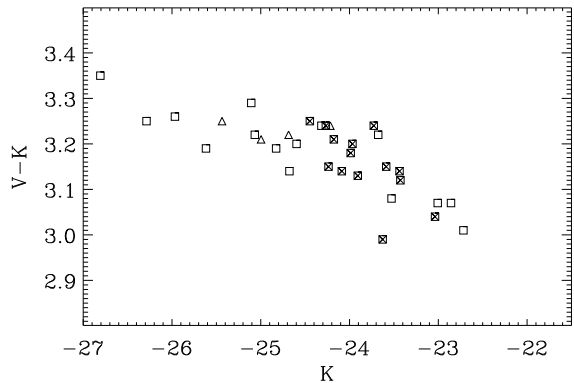}
\caption{The optical/NIR colour-magnitude relation of the Coma 
cluster using the photometry of Bower, Lucey \& Ellis (1992). 
Coma ETGs are shown as squares whereas the passive ETGs in the 
Virgo cluster are shown as triangles.
Galaxies with a K$_{\rm s}$-[16] excess are indicated with a cross.}
\label{fig:bowersel}
\end{figure}

\subsection{The mid-infrared colour-magnitude diagram}

In Figure~\ref{fig:cm12} we show the mid-infrared 
colour-magnitude diagram of the Coma cluster. The  K$_{\rm s}$-[16]
colour is shown as measured within a fixed aperture of 12 arcsec.

The distribution of K$_{\rm s}$-[16] colour shows a steady reddening from the 
faintest magnitudes observed ($\rm K_{s} \sim -22$) to $\rm K_{s} \sim -24$.
In this magnitude range the dispersion in K$_s-[16]$ is close to a magnitude
with values as high as 2.2 near $\rm K_s\sim -24$. At brighter magnitudes the 
colours show little variation and low dispersion with $\rm K_s-[16]\simeq 1.3$. 

We also plot the position of four selected
{\sl passive} ETGs in the Virgo cluster,
selected from the sample of Bressan et al. (2006).
These galaxies have nuclear MIR spectra 
whose main characteristic is the presence of the silicate
bump above $10\;\rm \mu m$, indicating a
passively evolving stellar population.
These galaxies have also been observed in the IRS Peakup mode
to secure a link between photometric and spectroscopic observations.

The presence of stellar population gradients within ETGs
(Annibali et al. 2007), together with the existence of a general 
increase of the effective radius with the total magnitude of the galaxy,
could, however, introduce a spurious effect, since a fixed aperture measurement
would probe stellar populations in relatively  more central regions for brighter
galaxies.

To quantify this effect we repeated the measurement within a fixed
{\sl relative} aperture of radius $\sim$2 R$_e$, where $R_e$ is the effective 
radius of the galaxy.
The results are shown in fig.~\ref{fig:re2}, though
the lack of measured H-band effective radii introduces a cut at magnitudes
fainter than K$_{\rm s}\sim -23$. 
In this figure the flattening at magnitudes brighter than
$\rm K_{s} \sim -24$ becomes more evident and actually the slope
of the colour-magnitude relation 
is reversed for the brightest Coma galaxies, while there are only
small differences at magnitudes fainter than $\rm K_{s} \sim -24$.

The position of the four passive Virgo galaxies in fig.~\ref{fig:cm12}
and that of the Coma galaxies brighter than K$_{\rm s}\sim -24$
define a tight relation that can be extended down to the faintest
magnitudes. 
By analogy with optical/NIR colour-magnitude relations and because
of the presence of the four Virgo ETGs
we argue that this line  represents the MIR colour-magnitude relation of 
\emph{passively evolving}, ETGs in the Coma cluster. ETGs significantly 
above this line have an ``excess'' of $16\;\rm \mu m$
emission with respect to the passive sequence.

The presence of a population of galaxies which appear anomalously
red in the mid-infrared colour-magnitude diagram is not mirrored in
the optical/near-infrared, fig.~\ref{fig:bowersel}. The V-K colour
shows a steady increase towards brighter magnitudes with a slight
flattening for objects brighter than $K\sim -24$. This, in fact, is
very similar to what we identify above as the \emph{passive} colour
magnitude relation in the mid- infrared. The implications of this
difference are discussed further below.

\section{Discussion}

From an extrapolation of the NIR stellar photospheric continuum into the 
mid-infrared Bressan et al. (2006) have estimated 
that the $16\;\rm\mu m$ emission 
in passive, ETGs has 
approximately equal contributions from stellar photospheres and dust emission.
The most likely origin of the latter emission is from dusty circumstellar
envelopes around evolved stars, 
such as the mass losing Asymptotic Giant Branch (AGB) stars. The exact
proportions depend on parameters such as the age and metallicity of the
stellar population. Because the photospheric contribution shows little variation
for a fixed K-band magnitude, the mid-infrared colour-magnitude relation should
depend strongly on the population of dusty AGB stars as a function of galaxy
luminosity. In this scenario, those ETGs that show an excess of 
$16\;\rm\mu m$ emission have the largest luminosity-weighted dusty AGB population. 
The evolution of this stellar population should therefore be critical to the 
interpretation of the mid-infrared colour-magnitude diagram.

Another possible origin of an excess of emission at $16\;\rm\mu m$
could be the contribution of a
central AGN,  as seen in the spectrum of NGC~4486 (Bressan et al. 2006).
However, there is no evidence of broad emission lines in the 
available SDSS spectra of those galaxies which lie well above the
`passive' colour-magnitude relation in fig.~\ref{fig:cm12}. 
There is also no evidence that these galaxies have smaller effective 
radii at $16\;\rm\mu m$ (as shown by NGC~4486, that appears
unresolved in the SL and LL IRS apertures).

\subsection{Influence of the inter-stellar environment}
\label{envelope}
Before considering the colour-magnitude relation in terms of stellar evolution, we
first consider the influences that the inter-stellar environment might have on the 
mid-infrared colours. 
In nearby AGB stars a significant fraction of the $16\;\rm\mu m$ flux
originates from extended, dusty, circumstellar envelopes.
It is  thus important to understand how
robust these envelopes are to different external perturbations, 
such as the ram-pressure of the interstellar medium (ISM) in the central regions
of massive ellipticals.

The circumstellar envelope of an AGB star expands at a typical velocity of
$\sim 10\;\rm km\,s^{-1}$ which, at the condensation radius, R$_c\sim$10$^{14}$cm (Bressan et al. 1998,
Granato \& Danese 1994) 
is of the order of the escape velocity. 
The MIR emission, f($\nu$), in AGB stars, has a broad
maximum between 10 and 20 $\mu$m suggesting a dust temperature 
T$_{dust}\sim 300\;\rm K$.
Inspection of our dusty circumstellar models shows that 
dust reaches this temperature only in a thin internal region of the envelope, 
whose size is a few tens of $R_c$. 
At $10\;\rm km\,s^{-1}$ the crossing time of this region is only 100~yr.
Thus, the MIR emission from AGB stars is a short lived phenomenon:
after about 100~yr, dust is already so cool that its contribution in 
the MIR spectral region becomes negligible.
However, in the central regions of an ETG,
the star moves at the dispersion velocity through the tenuous ISM, and we need to determine whether or not
the circumstellar envelope is able to maintain its structure in the inner 
few tens of $R_c$.

In order to estimate the radius at which the organized structure of the envelope 
is destroyed by its interaction with the ISM, we equate the kinetic energy density
in the circumstellar wind ($\sim  1/2 \rho_C v_\infty^2$) 
to the kinetic energy density of the 
ISM, due to the stellar dispersion velocity, ($\sim 1/2 \rho_I * \sigma_g^2$).

Using a stationary wind the circumstellar gas density is 
\begin{equation}
\rho_C = \frac{\dot{M}}{4\pi r^2 v_\infty}
\end{equation}
while the ISM density is evaluated from the mass that has been 
already lost by stars
\begin{equation}
\rho_I= \frac{d^2M}{dt dM_*} \times t \times \rho_*   \simeq  10^{-12} \times t \times  \rho_*
\end{equation}
In the second equation $d^2M/dtdM_*\simeq$$10^{-12} yr^{-1}$ is
the typical gas deposition rate per unit time and per unit stellar mass
from the old stellar population. This is obtained from integration of the mass loss rate along 
our isochrones of old stellar populations (see also Tinsley 1973). The average stellar density, 
$\rho_*$, is evaluated from deprojection of the  De Vaucouleurs law (Young 1976) and
depends on the assumed position within the galaxy. 

Applying the condition of virial equilibrium and the observed relation between
 the total galaxy mass and the velocity dispersion (Cappellari et al. 2006) we find
\begin{equation}
r/r_c \simeq  200 \sqrt{\frac{\dot{M_8}v_{10}}{L_3}
\frac{1}{\sigma_3^{1.78}\rho_{01} t_{\rm Gyr}}}
\end{equation}
In this equation, r$_c$ is the condensation radius, $\sim 1.66\times 
10^{12}~\sqrt{L/L_\odot}$, for a silicate dust mixture
(for graphite r$_c$ is about half this value and r/r$_c$ must be doubled);
$\dot{M_8}$ is the stellar mass loss rate in 10$^{-8}\rm M_\odot/yr$; $v_{10}$ is 
the wind velocity in units of $10\;\rm km\,s^{-1}$;
$L_3$ is the luminosity in units of 10$^3L_\odot$;   $\sigma_3$ is the stellar velocity
dispersion in units of $300\;\rm km\,s^{-1}$; r$_e$ is the effective radius of the galaxy;
$\rho_{01}$  is the dimensionless average stellar density
with respect to its value at $r/r_e=0.1$ ($\rho_{01}$ =10.5, Young, 1976);
M$_{12}$ is the galaxy mass in units of 10$^{12}M_\odot$ and 
t$_{\rm Gyr}$ is the time in Gyr.

Thus, only for very low mass-loss rates (10$^{-10}\rm M_\odot/yr$)
or in the very central regions of a galaxy - $\rho(r/r_e=0.01)=357$ -
may the structure of the wind be destroyed at a few condensation
radii. We may thus conclude that, in all relevant cases, the circumstellar wind 
is not destroyed by its interaction with the environment
even in the cores of ETGs.

Once the circumstellar envelope  mixes with the ISM,
dust grains may be destroyed via sputtering by the hot ($\sim 10^7\;\rm K$) ISM gas.
This process, however, will not affect the MIR emission, because, as discussed above,
when the dust is released to the ISM it is already too cool to emit
in the MIR spectral region. 
Furthermore, the timescale for dust grain destruction via sputtering, even
in the centre of the Coma cluster, 
is $\sim 10^8\;\rm yr$ (Dwek, Rephaeli \&
Mather, 1990), much larger than the lifetime of the newly formed hot dust
within the circumstellar envelope ($\sim$100 yr).

In summary, we conclude that the MIR emission from dust grains in AGB 
stars is a short lived phenomenon ($\sim 100\;\rm yr$) happening in the inner, denser 
region of the circumstellar envelopes. This emission cannot easily 
be modified by interaction with the environment, so that the 
environment does not have a \emph{direct} influence on the mid-infrared 
colour-magnitude relation.

\subsection{Stellar Evolution}

As it appears that the form of the MIR colour-magnitude relation is not 
influenced directly by some particular environmental effect, it must be
explained via the properties of the underlying stellar populations.

\begin{figure}
\centering
\includegraphics[clip,width=6cm,angle=90]{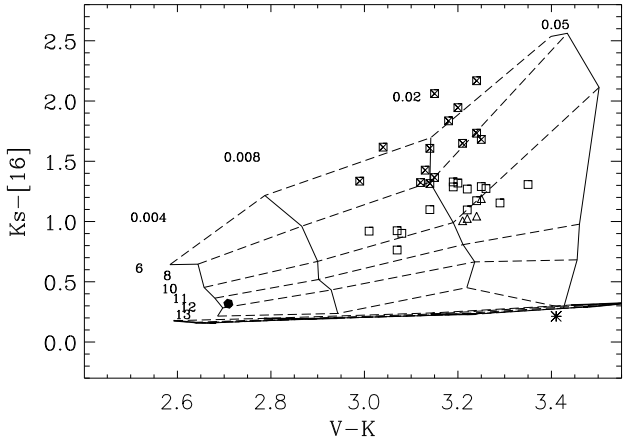}
 \caption{Optical NIR MIR colour-colour diagram. The horizontal 
 axis is the V-K colour from the precision photometry of Bower, 
Lucey \& Ellis (1992) while the vertical axis is the K$_{\rm s}$-[16] 
colour measured within fixed apertures of $12^{\prime\prime}$. 
 Squares are Coma ETGs and triangles are the four
  selected passive ETGs in the Virgo cluster. Galaxies with a K$_{\rm s}$-[16] 
excess are indicated with a cross. SSP models for
 the indicated ages and metallicity are computed following the scheme 
 of Bressan et al. (1998). Models without dusty AGB envelopes 
 collapse to the almost horizontal dashed lines at the bottom of the diagram. 
The dotted line represents the locus of the Kurucz model atmosphere 
 with low metallicity
 (the effect of metallicity is very small however) and the asterisk is a {\sc comarcs} 
 low temperature atmosphere model as calculated by Aringer et al. (2008, in prep.) 
The position of the globular cluster, 47-Tuc, is indicated by the solid hexagon in the lower left.}
\label{fig:colourcolour}
\end{figure}

\begin{figure}
\centering
\includegraphics[width=8.6cm]{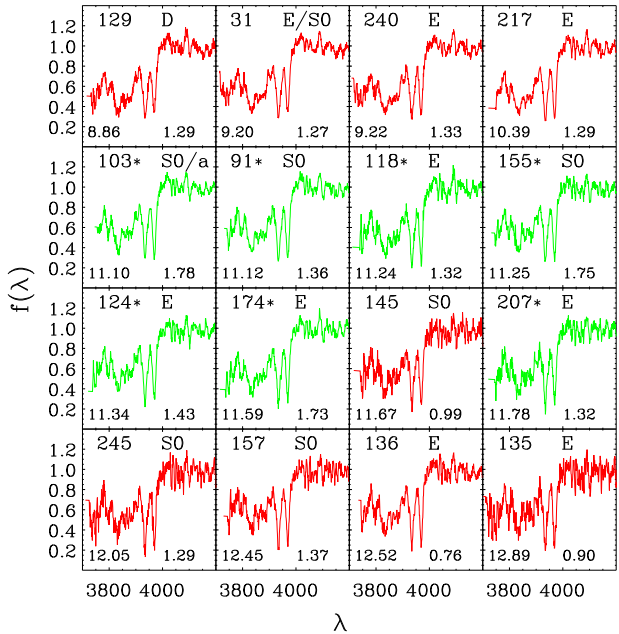}
  \caption{Optical SDSS spectra of selected Coma 
  cluster galaxies. Galaxies with a K$_{\rm s}$-[16] excess are shown in green and 
are indicated with an asterisk.}
\label{fig:sdss}
\end{figure}

\begin{figure}
\includegraphics[clip,width=8cm]{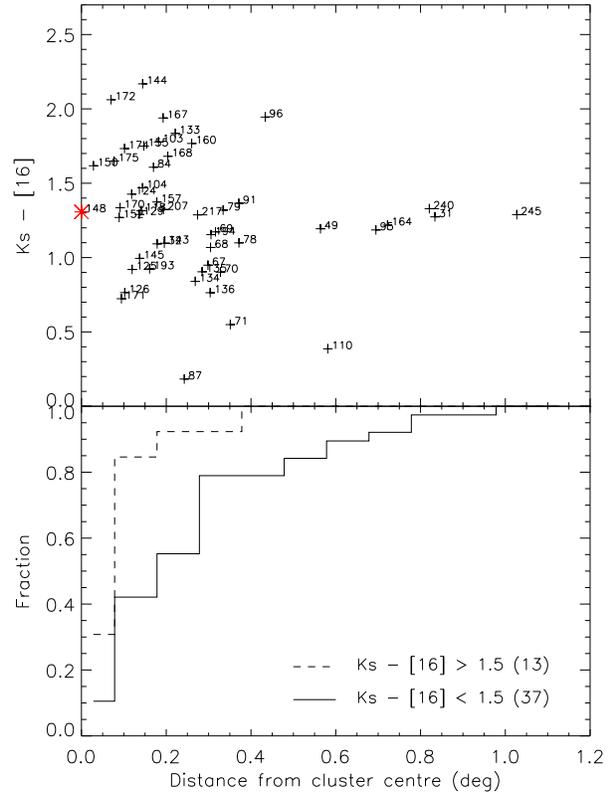}
 \caption{K$_{\rm s}$-[16] colour as a function of the projected distance from 
the cluster centre, defined as the CD galaxy D148 (asterisk on y-axis). The bottom 
panel shows the cumulative distribution for ETGs with K$_{\rm s}$-[16] $>1.5$ and 
K$_{\rm s}$-[16] $< 1.5$. The number of objects in each sub-set is indicated in
parentheses}
\label{fig:ccentre}
\end{figure}

In Figure \ref{fig:colourcolour} we compare our observations with
the predictions for the integrated colours of simple stellar populations (SSPs).

As discussed in Bressan et al. (1998) both the integrated optical and MIR
colours of SSPs suffer from the degeneracy between age 
and metallicity, but in the opposite sense.
While optical features (colours or narrow band indices) of the same strength
can be obtained by an {\sl anti-correlated} variation
of age and metallicity, producing the typical banana-shaped
solutions in the age-metallicity diagrams (Annibali et al. 2007),
an equal intensity of the 10$\mu$m bump may be obtained by
a {\sl correlated} variation of the age and metallicity.

For this reason, the effects of age and metallicity variations
are best separated in a mixed optical-NIR-MIR colour-colour plot,
such as that in Fig.~\ref{fig:colourcolour}.

In Fig.~\ref{fig:colourcolour} the horizontal axis is the V-K colour obtained 
from the precision photometry of Bower, Lucey \& Ellis (1992),
while the vertical axis is our measured K$_{\rm s}$-[16] colour. 
The K-band magnitudes on the two axes therefore refer to slightly different 
central filter wavelengths and apertures.\footnote{The photometry of Bower, Lucey \& 
Ellis was measured in an aperture of radius 8\farcs5 and corrected to a radius of 
5\farcs5. The aperture correction applied was 0.03 mag for V-K. Assuming a correction 
of similar magnitude would be required to correct to our radius of $12^{\prime\prime}$, 
the V-K values should shift by -0.03 mag. This would not alter any of our interpretation.}  
Though the V-K colours of Bower, Lucey \& Ellis are
provided only for a sub-sample of our galaxies (30 out of 50 objects),
we prefer to use their values to preserve the homogeneity of the data.
Squares are the Coma ETGs while the triangles refer to the four Virgo galaxies. 

New SSP models, for the indicated age and metallicity,
have been computed following the scheme of Bressan et al. (1998). 
The only difference with respect to these previous models is 
a minor revision of the mass loss rate during the AGB phase,
introduced by the calibration with new V-K integrated colours
of populous star clusters in the Large Magellanic Cloud,
and the adoption of the new empirical library of stellar spectra 
(S{\'a}nchez-Bl{\'a}zquez et al., 2006b).

A few caveats concerning models should be kept in mind because they are
relevant to the present investigation.

First of all, it is worth recalling that in Bressan et al. (1998)
the effects of possible circumstellar dust in the red giant branch (RGB) 
phase (i.e. before central helium burning in the horizontal branch in the case of 
old populations) have not been considered.
Recent claims, that RGB stars also show a significant
MIR excess not dependent on their luminosity (Origlia et 
al. 2007), should be regarded with great care
because of possible strong crowding effects (Boyer et al. 2008).

Furthermore, old stellar populations in the galaxies we are considering
could reach super-solar metallicity and,
as such, may not be well represented by
those in our own galaxy. Very little is actually known on the efficiency of the 
mass-loss rate at  high metallicity (Carraro et al. 1996).
However, van Loon et al. (2008)
do not find evidence  of a MIR excess in the RGB stars of NGC~6791,
a super metal-rich globular cluster.
Also, NGC~4649 and NGC~1399, the two ETGs with the largest
UV upturn and high Mg2 index (Bertola et al. 1995),
that have been interpreted as the signature of very metal-rich stars
(Bressan, Chiosi \& Fagotto 1994), have Spitzer
IRS spectra similar to those observed in our passive
Virgo ETGs (Bressan et al. in preparation).
Thus, there seems to be no evidence that metallicity has a
direct effect on mass loss.

As already anticipated, variations of age and metallicity are very well 
separated in this colour-colour diagram, especially at high metallicity, 
with iso-metallicity lines being almost horizontal and coeval SSPs
moving on almost vertical lines. In contrast, models without dusty AGB 
envelopes collapse to the almost horizontal dashed lines at the bottom of 
the diagram, independent of their ages. The effects of dust disappear at 
ages larger than 13 Gyr, where the  AGB phase is practically absent.

We have extended the photospheric stellar spectra of MILES
in the NIR/MIR region by means of matched NEXTGEN models (Hauschildt et al. 
1999).
In addition, the effects of dusty envelopes have been included
following the prescriptions described in Bressan et al. (1998).
Thus, the bottom horizontal dashed lines (models without
dusty AGB envelopes) superimpose on single NEXTGEN spectra
at varying effective temperature. The effect of varying the
metallicity is almost negligible in the 
K$_{\rm s}$-[16] colour. To illustrate the difference
introduced by adopting different stellar spectra
we have marked with an asterisk the position of a {\sc comarcs} 
 low temperature atmosphere model as calculated by Aringer et al. 
(2008, in prep.). Finally, the hexagon towards the bottom left
of the diagram (age $\sim$ 12 Gyr and Z $\sim$ 0.004),
marks the position of the globular cluster 47 Tuc
whose semi-empirical integrated colours have been
derived by Bressan et al. (2006b).

Fig.~\ref{fig:colourcolour} shows that galaxies that follow the MIR colour-magnitude
relation are almost coeval, with an age of about 10 Gyr and a metallicity that grows 
by a factor of about two. It is interesting to note that the four Virgo galaxies
are as old as the oldest Coma galaxies. None of the galaxies appear to be as old as 
the globular cluster 47 Tuc.

Galaxies with a K$_{\rm s}$-[16] excess, cluster around the lines of 8 Gyr and 6 Gyr
thus appearing younger by about 2-4 Gyr.

In order to clarify the different sensitivity to stellar populations
of optical and MIR colours, we have reproduced, in Fig.~\ref{fig:bowersel},
the optical/NIR colour-magnitude relation for the galaxies shown in 
Fig.~\ref{fig:colourcolour}. Galaxies with a K$_{\rm s}$-[16] excess
(i.e. clustering around and above 8 Gyr in Fig.~\ref{fig:colourcolour})
are plotted with a cross.
Fig.~\ref{fig:bowersel} shows that in spite of a well detected K$_{\rm s}$-[16] 
excess, the latter galaxies lie on the optical/NIR colour-magnitude relation. 
Moreover, a significant fraction of these galaxies are classified as ellipticals.

\subsection{Indirect influence of cluster environment on the K$_{\rm s}$-[16] colour ?}

The fraction of  galaxies showing a K$_{\rm s}$-[16] colour excess with
respect to the oldest systems, in Fig.~\ref{fig:colourcolour},  is about 48\%.
If we consider the total sample (Fig.~\ref{fig:cm12})
and add four more galaxies with a clear K$_{\rm s}$-[16] excess
with respect to the CM relation, the fraction is $32 \pm 8\%$.

A possible interpretation, discussed in the previous section,
is that these systems have average younger ages, since a younger stellar 
population is expected to have a redder MIR emission by virtue of the 
larger fraction of dusty AGB stars. However, galaxies are complex systems 
that may not be well represented by simple stellar populations and caution 
must be used in the interpretation of the K$_{\rm s}$-[16] colour excess. It does not
necessarily point to a younger global age of a passively evolving system.

For example, in the case of the Virgo ETG, 
NGC~4435, Panuzzo et al (2007) have shown that the MIR emission is dominated 
by a nuclear starburst which is practically invisible in the optical and NIR.
Its IRS low resolution spectra are dominated by PAH, atomic and molecular
emission features that are typical of star forming galaxies. At 16$\mu$m the 
emission is not resolved, indicating that the emission arises from the nucleus; 
which indeed appears as a dusty disk in HST images. In this rejuvenation episode 
(about 200 Myr ago) less than 1\% of the total galaxy mass was formed.
From Panuzzo et al. (2007) we estimate, for NGC~4435, a K$_{\rm s}$-[16] colour 
$\simeq$2.8 mag in the central 5 arcsec radius aperture, and a total K$_{\rm s}$-[16] 
colour $\simeq$2.1 mag and total V-K $\simeq$3.1. In Fig.~\ref{fig:colourcolour}, 
the latter values would put NGC~4435 in the region around 6 Myr and Z=0.02.
Indeed, NGC~4435 was selected for IRS observations as one of the galaxies
that make up the Virgo colour-magnitude relation. From the analysis of this sample
(18 galaxies) Bressan et al. (2006) concluded that the fraction of early-type 
galaxies that show signatures of recent activity is 17\%. This figure is 
significantly lower than the fraction of Coma ETGs with a 
K$_{\rm s}$-[16] colour excess.

We have also extracted, from data release 6 of the Sloan Digital Sky Survey (SDSS)
the available spectra (16 out of 50) of our Coma galaxy sample. In Fig.~\ref{fig:sdss} 
we show the $3700\rm \AA-4200\rm \AA$  spectral region, around the CaII lines. Galaxies are 
sorted by decreasing brightness and green curves represent objects with 
a K$_{\rm s}$-[16] colour excess. In three objects (D103, D118 and D155) with a K$_{\rm s}$-[16] excess, 
the CaII H line is as deep as the K line, indicating  contamination by A type stars,
i.e. a recent rejuvenation episode. In the other galaxies the CaII doublet
appears normal. We recall, that for solar metallicity, an inversion of the 
CaII doublet starts appearing only at ages younger than 2 Gyr (Longhetti et al., 1999). 

Although we see in Fig.~\ref{fig:ccentre} that the K$_{\rm s}$-[16] colour is
uncorrelated with the position in the cluster we note that if we identify
those galaxies with an `excess' of $16\;\rm\mu m$ emission as those with
$\rm K_s - [16] > 1.5$ (see fig.~ \ref{fig:cm12}) then the mean projected
distance from the cluster centre of this population is smaller than for
those galaxies with $\rm K_s - [16] < 1.5$. The mean projected distance
from the cluster centre for the  $16\;\rm\mu m$ excess galaxies is
$0.17^{\circ}$ whereas that for the others is $0.32^{\circ}$ (the KS-probability 
that the 2 distributions are the same is 1.6\%). Because it
seems that the cluster environment does not directly influence the mid-infrared 
colour-magnitude diagram this suggests that the star formation
history has been modified by the environment; which in turn influences
the galaxy colours. The sense of this influence is that galaxies with
redder K$_{\rm s}$-[16] colour (younger luminosity weighted stellar ages, 
fig.~\ref{fig:colourcolour}) are preferentially found at small cluster-centric
radii.

Miller, Neal \& Owen (2002) have investigated the distribution of both
AGN and star-forming galaxies as a function of cluster-centric distance.
They find that dusty star-forming galaxies have a more centrally
concentrated distribution than normal star-forming galaxies, with AGN
being more centrally concentrated than both. 
Though there is no evidence of emission lines in the SDSS
spectra, we cannot exclude that the MIR excess is due
to an obscured nuclear starburst like that found in NGC~4435.
A possible way to check this possibility is by means of existing IRAC 4.5$\mu$m
and 8$\mu$m observations (sampling the strength of the 7.7 PAH feature) and 
MIPS 24$\mu$m (sampling the presence of hot dust). This work is underway.

\subsection{Implications}

The MIR colour-magnitude relation indicates that the oldest Coma early-type 
galaxies define a sequence of increasing metallicity at constant age 
($Z\simeq 0.01-0.03$, $t\simeq 10\;\rm Gyr$).

While this conclusion has already been suggested by the analysis of the 
optical/NIR colour-magnitude relation, it is only with the use of the 
complementary MIR observations that the alternative solution (invoking 
a synchronization of the epoch of formation with total brightness,
see Bower, Lucey \& Ellis, 1992) can be excluded.

Indeed, our analysis of the mixed optical-NIR-MIR colour-colour diagram 
clearly shows that some of the galaxies that populate the optical 
colour-magnitude relation may have average ages that are even 40\% younger
than the old sequence. In those galaxies, however, a larger metallicity 
may redden the colours and compensate for the younger age in an optical diagram.
This illustrates the importance of the MIR colour in breaking the 
age-metallicity degeneracy.

Our conclusion is in very good agreement with the analysis of narrow band indices of 
$\sim 4000$ ETGs carefully selected from the SDSS survey
(Clemens et al. 2006) and recently extended to $\sim 14000$ objects
with the DR6 catalogue (Clemens et al. 2008 in prep.). These authors find that the 
luminosity-weighted ages of the whole sample start young at a low velocity
dispersion, rise until $\sigma\sim 2.32$ and flatten at larger $\sigma$. In contrast, 
the metallicity shows a continuous increase with increasing $\sigma$.

Our result, that a large fraction of ETGs in Coma 
populate a genuine old and coeval sequence, contrasts somewhat with 
the recent study of ETGs in
the Coma cluster by
Trager et al. (2008). Based on optical spectra
these authors also find that, in spite of a wide range in mass,
10/12 objects
are consistent with the hypothesis of a coeval
population. However, the age they find for this population,
$5.2\pm 0.2\;\rm Gyr$, is significantly younger
than our value.

We notice that this discrepancy is unlikely to be caused by
a significant offset between ages derived with the two methods.
Indeed, there are four objects in common, for which Trager et al. 
find the following ages:  
D129 (GMP3329) 7.9 Gyr, D157 (GMP3484)  7.8 Gyr, 
D155 (GMP3367) 4.5 Gyr and  
D133 (GMP3639) 3 Gyr.

D129 and D157, the two old objects, fall on our MIR colour-magnitude relation, for which
we derive only sligthly older ages. On the other hand D155 and D133, the two young objects,
lie well above this relation and, given their K$_{\rm s}$-[16] colour we would conclude that
they are possibly affected by some rejuvenation episode.

Considering that, contrary to our analysis, optical studies
are affected by the age-metallicity degeneracy causing 
(large) uncertainties that are difficult to quantify,
we believe that there is good agreement for the few objects in common.

It is more difficult to explain why 
we find that 68\% of the galaxies in our sample populate the old
sequence, while Trager et al. find that only 25\% are old.
If their sample is not biased toward young objects and
given the sensitivity of the K$_{\rm s}$-[16] colour
to rejuvenation events, we suspect that the age-metallicity 
degeneracy is still to blame.

Bregman et al. (2006) find a very similar discrepancy between the age 
of ETGs determined by optical line indices and mid-infrared Spitzer spectra
that include the silicate emission feature near $10\;\rm \mu m$. They also find 
mean ages of $\sim 10\;\rm Gyr$ using mid-infrared data. 

It is also interesting to note that the upper 2.5 mag of the MIR 
colour-magnitude relation is quite narrow. On the contrary, 
in the lower 2.5 mag, below M$_K\sim -24.5$, there is a significant 
dispersion. This shows that the rejuvenation events (of any kind)
have not affected the most massive ETGs.

\section{Conclusions}

We present $16\;\rm \mu m$, Spitzer-IRS, blue peakup images of a sample
of 50 ETGs in the Coma cluster. We compare these with archival IRAC
images at  $4.5\;\rm \mu m$ and 2MASS, $K_s$ band images at
$2.2\;\rm \mu m$.

We make the following conclusions.

\begin{itemize}

\item{Our IRS blue peakup images show no evidence that the $16\;\rm \mu m$ 
emission is anything other than stellar in origin}

\item{The region within a dusty AGB star envelope where dust is hot
enough to emit strongly at $16\;\rm \mu m$ is not vulnerable to
environmental effects such as the ram-pressure of an ISM wind or dust
sputtering by hot gas. Dust grains only spend $\sim 100\;\rm yr$ near
$\sim 300\;\rm K$ and this region is sufficiently dense to survive
against ram-pressure disruption.}

\item{We identify the mid-infrared colour-magnitude relation of 
passively evolving ETGs as the lower envelope of the
galaxy distribution in the K$_s$ - [16] vs K$_s$ plane; that is, the
minimum value of K$_s$ - [16] at a given K$_s$ magnitude. $\sim$68\% of the 
galaxies in our sample lie on this colour-magnitude 
relation. These galaxies cannot have had any episode of star
formation accounting for more than $\sim 1\%$ of the total stellar mass
within the last few Gyr. These are genuinely `passively evolving'
objects. The remaining objects are either significantly younger than 10 Gyr
or have undergone a rejuvenation event in the recent past. This result is 
at odds with the most recent estimate of the fraction of old objects based 
on optical spectroscopy (Trager et al. 2008).}

\item{We construct the mixed optical-NIR-MIR two colour diagram and, 
by means of updated simple stellar population models, we show that
the addition of mid-infrared data allows a much better separation of the 
effects of age and metallicity, which are rather degenerate in either the 
optical or mid-infrared when taken in isolation. In this plane galaxies populating the
colour-magnitude relation trace a sequence of \emph{varying metallicity at 
approximately constant age}. Although this conclusion was already consistent 
with the optical-NIR colour-magnitude relation, a correlation between age 
and metallicity could equally well explain the relation. Indeed, comparison 
with the optical-NIR colours shows that \emph{a number of galaxies that lie on the
optical-NIR relation are significantly displaced from the mid-infrared
relation}, with redder K$_s$ - [16] colours. The mid-infrared colour-magnitude 
diagram therefore shows a sequence of metallicity for old, passive galaxies, with 
younger objects displaced towards redder K$_s$ - [16] colours.}

\item{The oldest elliptical galaxies in our sample have luminosity
weighted, mean stellar ages of 10.5 Gyr and metallicities within a factor
of two of the solar value. No galaxy in our sample is as old or metal
poor as the globular cluster 47 Tuc.}

\item{Although the addition of the K$_s$ - [16] colour allows us to
identify objects with significantly younger, luminosity weighted, mean
stellar ages, we cannot distinguish between genuinely `young' objects and
those that have undergone a minor rejuvenation event. However, given that
even a period of recent (last few Gyr) star formation that accounts for
less than 1\% of the total stellar mass will shift a galaxy off the mid-infrared 
colour-magnitude relation, the latter option seems far more
likely. Unambiguous resolution of this issue will require the infrared spectroscopic
capabilities of future space observatories.}

\item{There is evidence that those galaxies that have an excess in the 
K$_s$ - [16] colour are found preferentially at smaller cluster-centric
radii. As the interaction between the dusty AGB star envelopes and the
ISM does not directly effect the $16\;\rm \mu m$ emission, the excess may
be caused by ``rejuvenation'' episodes induced by the cluster environment.}

\end{itemize}

\section*{Acknowledgments}
We acknowledge a financial contribution from contract ASI-INAF I/016/07/0.

This work is based on observations made with the Spitzer Space Telescope, which
is operated by the JPL, Caltech under a contract with NASA.

We make use of data products from the Two Micron All Sky
Survey, which is a joint project of the University of Massachusetts and
the Infrared Processing and Analysis Center/California Institute of
Technology, funded by the National Aeronautics and Space Administration
and the National Science Foundation. 

This research has made use of the GOLD Mine Database.

Funding for the SDSS and SDSS-II has been provided by the Alfred P. Sloan
Foundation, the Participating Institutions, the National Science
Foundation, the U.S. Department of Energy, the National Aeronautics and
Space Administration, the Japanese Monbukagakusho, the Max Planck
Society, and the Higher Education Funding Council for England. The SDSS
Web Site is http://www.sdss.org/.

The SDSS is managed by the Astrophysical Research Consortium for the
Participating Institutions. The Participating Institutions are the
American Museum of Natural History, Astrophysical Institute Potsdam,
University of Basel, University of Cambridge, Case Western Reserve
University, University of Chicago, Drexel University, Fermilab, the
Institute for Advanced Study, the Japan Participation Group, Johns
Hopkins University, the Joint Institute for Nuclear Astrophysics, the
Kavli Institute for Particle Astrophysics and Cosmology, the Korean
Scientist Group, the Chinese Academy of Sciences (LAMOST), Los Alamos
National Laboratory, the Max-Planck-Institute for Astronomy (MPIA), the
Max-Planck-Institute for Astrophysics (MPA), New Mexico State University,
Ohio State University, University of Pittsburgh, University of
Portsmouth, Princeton University, the United States Naval Observatory,
and the University of Washington.

We thank P. Marigo, L. Girardi and A. Renzini for useful discussions.

\bsp

\label{lastpage}


\begin{thebibliography}{}

\bibitem[]{}Annibali F., Bressan A., Rampazzo R., Zeilinger W.~W., Danese L., 2007, A\&A, 463, 455 
\bibitem[]{}Aringer B. et al. 2008, A\&A in prep.
\bibitem[]{}Athey A., Bregman J., Bregman J., Temi P., Sauvage M., 2002, ApJ, 571, 272 
\bibitem[]{}Bernardi M., Sheth R.~K., Nichol R.~C., Schneider D.~P., Brinkmann J., 2005, AJ, 129, 61 
\bibitem[]{}Bertola, F., Bressan, A., Burstein, D., Buson, L.~M., Chiosi, C., \& di Serego Alighieri, S.\ 1995, ApJ, 438, 680 
\bibitem[]{}Bower R.~G., Lucey J.~R., Ellis R.~S., 1992, MNRAS, 254, 589 
\bibitem[]{}Boyer, M.~L., McDonald, I., van Loon, J.~T., Woodward, C.~E., Gehrz, R.~D., Evans, A., \& Dupree, A.~K., 2008, AJ, 135, 1395
\bibitem[]{}Bregman J.~N., Temi P., Bregman J.~D., 2006, ApJ, 647, 265
\bibitem[]{}Bressan, A., Chiosi, C., \& Fagotto, F., 1994, ApJS, 94, 63
\bibitem[]{}Bressan A., et al., 2006, ApJ, 639, L55 
\bibitem[]{}Bressan A., Granato G.-L., Silva L., 1998, A\&A, 332, 135  
\bibitem[]{}Cappellari M., Bacon R., Bureau M., Damen M.~C., Davies R.~L., de Zeeuw P.~T., Emsellem E., Falcon-Barroso J., Krajnovic D., Kuntschner H., McDermid R.~M., Peletier R.~F., Sarzi M., van den Bosch R.~C.~E., van de Ven G., 2006, MNRAS, 366, 1126\\
\bibitem[]{}Carraro, G., Girardi, L., Bressan, A., \& Chiosi, C.\ 1996, A\&A, 305, 849 
\bibitem[]{}Clemens M.~S., Bressan A., Nikolic B., Alexander P., Annibali F., Rampazzo R., 2006, MNRAS, 370, 702 
\bibitem[]{}Dressler, A., 1980, ApJS, 42, 565 
\bibitem[]{}Dwek E., Rephaeli Y., Mather J.C., 1990, ApJ, 350, 104
\bibitem[]{}Gavazzi G., Franzetti P., Scodeggio M., Boselli A., Pierini D., 2000, A\&A, 361, 863 
\bibitem[]{}Gavazzi G., Boselli A., Donati A., Franzetti P., Scodeggio M., 2003, A\&A, 400, 451
\bibitem[]{}Gledhill T.~M., Yates J.~A., 2003, MNRAS, 343, 880 
\bibitem[]{} Granato, G.~L., \& Danese, L.\ 1994, MNRAS, 268, 235
\bibitem[]{}Hauschildt, P.~H., Allard, F., Ferguson, J., Baron, E., \& Alexander, D.~R., 1999, ApJ, 525, 871
\bibitem[]{}Landsman W.B., 1995, ASP Conference Series, 77, 437
\bibitem[]{}Lebzelter, T., Posch, T., Hinkle, K., Wood, P.~R., \& Bouwman, J.\ 2006, ApJ, 653, L145 
\bibitem[]{}Leeuw L.~L., Sansom A.~E., Robson E.~I., Haas M., Kuno N., 2004, ApJ, 612, 837 
\bibitem[]{}Longhetti, M., Bressan, A., Chiosi, C., \& Rampazzo, R., 1999, A\&A, 345, 419
\bibitem[]{}Marleau F.~R., et al., 2006, ApJ, 646, 929 
\bibitem[]{}Miller, Neal A.; Owen, Frazer N., 2002, AJ, 124, 2453
\bibitem[]{}Origlia, L., Rood, R.~T., Fabbri, S., Ferraro, F.~R., Fusi Pecci, F.,\& Rich, R.~M., 2007, ApJ, 667, L85
\bibitem[]{}Panuzzo P., et al., 2007, ApJ, 656, 206
\bibitem[]{}Renzini, A., 2006, ARA\&A, 44, 141
\bibitem[]{}S{\'a}nchez-Bl{\'a}zquez P., Gorgas J., Cardiel N., Gonz{\'a}lez J.~J., 2006b, A\&A, 457, 809 
\bibitem[]{}S{\'a}nchez-Bl{\'a}zquez P., et al., 2006b, MNRAS, 371, 703
\bibitem[]{}Temi P., Brighenti F., Mathews W.~G., 2008, ApJ, 672, 244 
\bibitem[]{}Temi P., Brighenti F., Mathews W.~G., 2007, ApJ, 660, 1215 
\bibitem[]{}Thomas D., Maraston C., Bender R., Mendes de Oliveira C., 2005, ApJ, 621, 673 
\bibitem[]{} Tinsley, B.~M., 1973, ApJ,186, 35
\bibitem[]{}Trager S.~C., Faber S.~M., Dressler A., 2008, MNRAS, 386, 715
\bibitem[]{}van Loon, J.~T., Boyer, M.~L., \& McDonald, I., 2008, ApJ, 680, L49
\bibitem[]{}Young P.~J., 1976, AJ, 81, 807       

\end{thebibliography}
\end{document}